\documentclass[final,1p,times]{elsarticle}

\nonstopmode
\usepackage[utf8]{inputenc}
\usepackage{amssymb,amsmath}
\usepackage{hyperref}
\usepackage{color}

\def\htot{$H^{\text{tot}}$}
\def\thetarf{$\phi_{\text{RF}}$}
\def\thetarfn[#1]{$\phi_{\text{RF}}=#1^{\circ}$}
\def\li7{$^7$Li}
\def\h1{$^1$H}

\def\bldE{\boldsymbol{E}}

\def\bldH{\boldsymbol{H}}
\def\bldn{\boldsymbol{n}}
\def\bldJ{\boldsymbol{J}}

\begin{document}

\begin{frontmatter}

\title{Visualizing Skin Effects in Conductors with MRI: \li7{} MRI Experiments
and Calculations}

\author[nyuchem] {Andrew J. Ilott}
\author[maglab]  {S. Chandrashekar}
\author[uiuc]    {Andreas Klöckner}
\author[sbu]     {Hee Jung Chang}
\author[sbu]     {Nicole M. Trease}
\author[sbu,camb]{Clare P. Grey}
\author[nyumath] {Leslie Greengard}
\author[nyuchem] {Alexej Jerschow\corref{cor1}}
\ead{alexej.jerschow@nyu.edu}

\address[nyuchem]
{Department of Chemistry, New York University, 100 Washington Square East, New York, NY 10003, USA.}
\address[maglab]
{National High Magnetic Field Laboratory and Florida State University, 1800 E. Paul Dirac Drive, Tallahassee, FL 32310, USA.}
\address[uiuc]
{Department of Computer Science, University of Illinois at Urbana-Champaign, Urbana, IL, 61801, USA.}
\address[sbu]
{Department of Chemistry, Stony Brook University, Stony Brook, NY 11794, USA.}
\address[camb]
{Department of Chemistry, University of Cambridge, Lensfield Road, Cambridge CB2 1EW, UK.}
\address[nyumath]
{Courant Institute of Mathematical Sciences, New York University, New York, NY 10012, USA.}
\cortext[cor1]{Corresponding author}

\begin{abstract}
While experiments on metals have been performed since 
the early days of NMR (and DNP), the use of bulk metal is normally avoided. 
Instead, often powders have been used in combination with low fields, 
so that skin depth effects could be neglected. Another complicating 
factor of acquiring NMR spectra or MRI images of bulk metal is the 
strong signal dependence on the orientation between the sample and the 
radio frequency (RF) coil, leading to non-intuitive image distortions and 
inaccurate quantification. 
Such factors are particularly important for NMR and MRI of batteries and other electrochemical devices. 
Here, we show results from a systematic study combining RF field 
calculations with experimental MRI of $^7$Li metal to
visualize skin depth effects directly and to analyze the RF field orientation 
effect on MRI of bulk metal. It is shown that a certain degree of  
selectivity can be achieved 
for particular faces of the metal, simply based on the orientation of the sample. 
By combining RF field calculations with bulk magnetic susceptibility calculations 
accurate NMR spectra can be obtained from first 
principles. Such analyses will become valuable in many 
applications involving battery systems, but also metals, in general.
\end{abstract}

\begin{keyword}
Magnetic Resonance Imaging \sep
Lithium batteries \sep
RF field calculations \sep
Susceptibility effects \sep
Skin effect
\end{keyword}

\end{frontmatter}


\section{Introduction}

It is well known that electromagnetic fields decay upon entering 
conducting regions \cite{jackson}. For good conductors, the fields can be shown to decay
exponentially with a characteristic distance, $\delta$, referred to as the skin depth,
\begin{equation}
 \delta = \sqrt {\frac{\rho}{\pi \mu_0 \mu_r \nu}},
\label{eq:sknDpth}
\end{equation}
where $\nu$ is the frequency of the field, $\mu_0$ the permeability of free space,
$\mu_r$ the relative permeability of the conductor and $\rho$ its resistivity.
At typical NMR frequencies (70--300~MHz for nuclei other than protons) $\delta$ is of the order of 10s of $\mu$m for good metal
conductors ($\rho\approx10^{-8}~\Omega$m).
The skin depth hence determines the limited region within which NMR-active nuclei can
be probed in a conducting sample.

In previous NMR studies of  metals, skin effects have been
typically avoided by using low magnetic fields and powdered samples to 
maximize sensitivity. This approach was appropriate for most early studies, 
where there was interest in measuring the Knight shifts and relaxation times of metal nuclei 
in order to probe the hyperfine interactions and gain insight into the 
electronic structure of metals \cite{gutowsky_1952,mcgarvey_nuclear_2004,narath_effects_1968,abati_1973,van_der_klink_2000}.
Indeed, there have been successful studies on powdered metals, both 
pure \cite{narath_effects_1968,andrew_nmr_1974,andrew_more_1971,el-hanany_1974,narath_nuclear_1967,narath_nuclear_1968}
and alloyed \cite{narath_nuclear_1967,matzkanin_1969,narath_nuclear_1968-1,bloembergen_1953,snodgrass_1971}.
Powdered metal conductors were also amongst the first samples proposed for dynamic nuclear polarization 
(DNP) transfer experiments \cite{overhauser_1953}.

Aside from the skin depth effects, which restrict sensitivity by limiting the 
effective sample size, samples containing bulk conducting metals are further 
afflicted by artifacts due to their interactions with the static and the RF  magnetic  fields used in MR experiments. Interactions with the static field can introduce large local $\boldsymbol{B}_0$ field inhomogeneities in the proximity 
of arbitrarily shaped, bulk metal that are a consequence of the typically large 
magnetic susceptibilities of many metals. The extent of these artifacts 
depends on the shape of the metal object and its relative orientation with 
respect to the external, $\boldsymbol{B}_0$ field. The second effect is due  to the 
conducting properties of the metal, which lead to the induction of surface 
currents when the sample is irradiated with an oscillating RF field \cite{bloembergen_1952}.
The induced currents modify the effect of the RF field at the surface of the metal 
and in the immediate vicinity of it, impacting the excitation and detection 
of the nuclear spins in these regions. The RF effect also depends 
on the sample shape and its orientation with respect to the applied RF,
as both of these factors dictate the form of the induced current. 

Traditionally, the above problems have been most strongly associated with reports of
artifacts in clinical MRI scanning originating from the presence of metallic 
implants \cite{hargreaves_2011}.
Recently, these effects have also become pertinent problems in the 
development of NMR and MRI techniques for the \textit{in situ} 
study of electrochemical 
devices. Here, the use of conducting components is clearly a requirement. 
Indeed, \textit{in situ} NMR \cite{bhattacharyya_situ_2010,blanc_situ_2013}
and MRI \cite{chandrashekar_7li_2012,klett_2012}
of functioning Li-metal batteries have given great insight into properties 
of the electrolyte and the metal electrodes during the charge cycles in such 
systems. The strong susceptibility effects associated with the regions 
surrounding the Li-metal have even proven to be advantageous for such studies, as they 
give rise to distinct and characteristic chemical shifts for surface microstructures and 
dendrites \cite{bhattacharyya_situ_2010,chandrashekar_7li_2012}.
As a result, 
chemical shift images (CSIs) not only revealed the location of such entities, 
but also some of their characteristics. In previous studies, qualitative rules have been followed to 
align the main face of the (square) conducting metal electrodes with the 
RF field direction in order to minimize RF field 
inhomogeneities, but the RF field dependence on orientation and shape were not known \cite{chandrashekar_7li_2012,britton_situ_2013,tang_2012}.

In this work, we combine RF field and bulk magnetic susceptibility 
calculations with experimental NMR and MRI results to clarify the 
role of the RF-field orientation, skin effects, and susceptibility 
shifts in determining the appearance of spectra and images of bulk 
metals and samples that contain them. 
The experiments and calculations are performed on 
\li7 metal, a moderately good conductor, but also a
highly relevant material in battery applications where
the insights provided should be of particular value. 
It is shown that all of the above effects can be accounted for explicitly, 
with an excellent agreement  between calculation and experiment. 
Furthermore, the calculation methods are robust and easily extended to 
more complicated systems. Insights from these should prove beneficial for 
future studies and aid in the development of promising 
novel techniques for bulk metal NMR and MRI. 

\begin{figure}
\begin{centering}
\includegraphics[width=6cm]{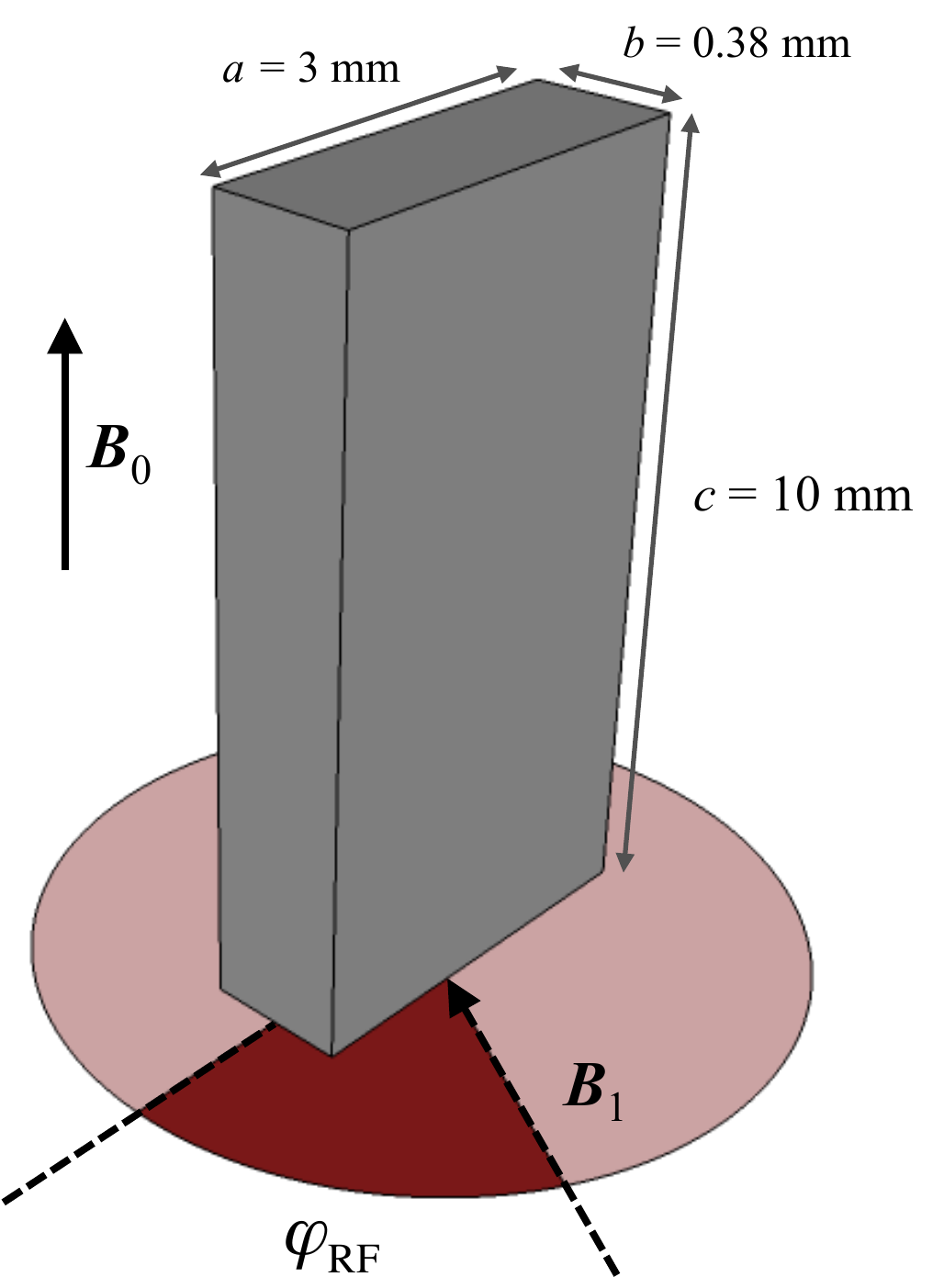}
\par\end{centering}
\caption{\label{fig:electrode}
Schematic of the cell used in the MRI experiments and 
calculations to represent the lithium metal electrode.
The $c$ axis of the rectangular strip
is aligned along the external magnetic field, $\boldsymbol{B}_0$.
The angle \thetarf{} specifies the relative orientation between the 
plane of the $ac$ metal surface with the magnetic field component of
the RF field. }
\end{figure}

\section{Methods}

\subsection{Magnetic resonance imaging.}

The sample consisted of a $\sim$10~x~3~x~0.38~mm strip of \li7{}
metal (Aldrich, 99.9\%) within a standard 5~mm o.d. NMR tube. 
Strips of glass microfiber
separator were placed on either side of the metal strip to center it in the tube
and to secure it tightly in place during the experiments. The sample was
assembled and sealed inside an argon glove box.

All MRI and NMR experiments were performed on a Bruker Ultrashield
9.4 T Avance I spectrometer containing a Bruker Micro2.5 gradient
assembly and operating at 155.51~MHz for \li7{}. A wide band HR-50
probe (Cryomagnet Systems Inc., Indianapolis) equipped with a 5~mm
saddle coil was used to collect all of the images. In this setup,
the sample is free to rotate in the probe and the probe free to rotate
inside the gradient stack, allowing the angle between the sample and
RF field direction, \thetarf{}, and between the sample
and principal gradient directions, $\phi_{\text{Gx}}$, to be set
independently. 

The images were acquired without slice selection using a spin echo
pulse sequence with 64 phase encoding steps incrementing the gradient
strength between $\pm$38.2~G~cm$^{-1}$ 
and applied for 750~$\mu$s with a 75~$\mu$s ramp time, giving a 
6.0~mm field of
view (FOV) and nominal 94~$\mu$m resolution. A 45.3~G~cm$^{-1}$
frequency encoding gradient was used during acquisition with a 30~kHz
spectral width, giving a FOV and nominal resolution of 4.0~mm and
31~$\mu$m. For each required orientation, \thetarf{},
the gradient orientation was kept constant, $\theta_{Gx}=0^{\circ}$,
so that the phase direction was parallel to the face of the metal strip,
and the frequency encoded direction perpendicular to it. The total
echo time, TE, and repetition time, TR, were 1.1~ms and 100~ms,
respectively. Pulses corresponding to a power of 5~kHz were used 
for excitation and refocusing. This same pulse power was also used 
to acquire a 1D NMR spectrum at each orientation. Chemical shifts 
were referenced to 1~M LiCl at 0.0~ppm. 

\subsection{Numerical Simulations.}

\subsubsection{Calculation of Electromagnetic Fields.}

For the numerical simulations, the electrodes were modeled as perfect
electric conductors scattering an incident plane wave with wave
number $k=100$~cm$^{-1}$, corresponding to an excitation at a wave length
of $\lambda=2\pi/k\approx 62 \text{ cm}$.
The incident wave was taken to be propagating along the $x$ axis and linearly
polarized along the $z$ axis.

The non-dimensionalized, time-harmonic, linear, isotropic Maxwell's
equations \cite{jackson}
\begin{align*}
   \nabla \times \bldH + ik\bldE &= 0\\
   \nabla \times \bldE - ik\bldH &= 0\\
   \nabla \cdot \bldE &=0\\
   \nabla \cdot \bldH &=0
\end{align*}
were used with perfect electric conductor (PEC) boundary conditions
\begin{equation}
  \bldn \times \bldE^{\text{tot}} = 0, \quad
  \bldn \cdot  \bldH^{\text{tot}} = 0
  \qquad \text{on the metal surface  }
  \label{eq:pec}
\end{equation}
where $\bldn{}$ is the unit normal to the metal surface.

These equations were then solved for the scattered field
\[
  \bldE = \bldE^{\text{tot}} - \bldE^{\text{inc}},\qquad
  \bldH = \bldH^{\text{tot}} - \bldH^{\text{inc}}.
\]
$\bldE^{\text{inc}}$ and $\bldH^{\text{inc}}$ here are defined
as the incident plane wave described above.

To obtain a numerical solution, the problem was recast as a variant of
the magnetic field integral equation \cite{maue_formulation_1949} with
surface charge and current densities $\rho$ and $\bldJ$ as unknowns
\cite{taskinen_current_2006,vico_mfie}
\begin{align}
    \label{eq:mfie}
    \bldJ/2 - \bldn\times (\nabla \times S_k \bldJ)
    &=
     \bldn\times \bldH^{\text{inc}}\\
    \label{eq:charge-eqn}
  \rho/2+\hat n \cdot \nabla S_k \rho
  &=
  \bldn\cdot \bldE^{\text{inc}} + ik(\bldn\cdot S_k \bldJ),
\end{align}
where $S_k$ is the single layer {\em scalar} potential operator for the
Helmholtz equation \cite{coltonkress}
\begin{equation}
  S_k \rho(\mathbf{x}):=\frac1{4\pi}\int_\Gamma
  \frac{e^{ik|\mathbf{x}-\mathbf{x'}|}}{|\mathbf{x}-\mathbf{x'}|}
  \rho(\mathbf{x'})ds_{x'} \, ,
  \label{eq:slp}
\end{equation}
and where $\mathbf{x}$ and $\mathbf{y}$ are coordinates in $\mathbb{R}^3$.

Note that the surface current density is a tangential vector field,
following from $\bldn \times \bldH^{\text{inc}}$,
so equations
\eqref{eq:mfie} and \eqref{eq:charge-eqn} form a (square) system of
Fredholm boundary integral equations of the second kind \cite{coltonkress}.

Equations \eqref{eq:mfie} and \eqref{eq:charge-eqn} were solved iteratively
using GMRES \cite{saad_gmres_1986}, and values for $\bldE$ and $\bldH$ were
recovered at all target locations from $\bldJ$ and $\rho$ according to the
representations
\begin{align*}
  \bldE(\mathbf{x})&= ikS_k \bldJ(\mathbf{x}) - \nabla S_k\rho(\mathbf{x}),\\
  \bldH(\mathbf{x})&= \nabla \times S_k \bldJ(\mathbf{x}).
\end{align*}
$\bldJ$ and $\rho$ were discretized as piecewise constants on a triangular
grid tessellating the surface of the electrode.  The integrals in the layer
potentials on each triangle were computed by ``singularity subtraction":
analytically for the static kernel $1/{|\mathbf{x}-\mathbf{y}|}$ and by high order Gaussian
quadrature for the smoother remainder $(e^{ik|\mathbf{x}-\mathbf{y}|}-1)/{|\mathbf{x}-\mathbf{y}|}$. The calculation
was accelerated by making use of the low frequency fast multipole method (FMM), 
described in \cite{cheng_wideband_2006}. 
FMM-accelerated integral equation methods permit the solution of
electromagnetic scattering problems in $O(N \log N)$ time, where $N$ denotes
the number of degrees of freedom in the discretization of the scatterer.
Unlike finite difference or finite element methods, integral equation
methods do not require a discretization
of the domain itself and impose radiation conditions at infinity by construction.
We made use of the open-source
\texttt{FMMLIB3D} library \cite{greengard_fmmlib_2012}.

Triangular surface meshes for the electrodes were generated
using the Gmsh mesh generator
\cite{geuzaine_gmsh_2009}. Care was taken to devote
a substantial amount of resolution to the singularities of charge
and current densities along edges and corners. Characteristic
lengths for the mesh were set to $3.5\cdot 10^{-7}\text{ m}$ at corners
and to $7\cdot 10^{-6} \text{ m}$ elsewhere.
%

Only the $x$ and $y$ components of the ${\boldsymbol{H}}^{\text{tot}}$ 
RF field are perpendicular 
to $\boldsymbol{B}_0$ and hence excite the precessing nuclear spins.
The nuclear magnetization nutates at a rate
that is proportional to the strength of the RF field at each of the $q$
triangles in the mesh,
\begin{equation}
  \omega_q = \frac{| H_{q}^{\text{tot}}(x) + H_{q}^{\text{tot}}(y)|}{2\pi}.
  \label{eq:omega}
\end{equation}
The variation of $\omega_q$ leads directly to a variation
in the detected signal at each triangle,
\begin{equation}
  F_{q}\propto\omega_q\sin\left(\omega_q\tau\right),
  \label{eq:trisignal}
\end{equation}
using the principle of reciprocity \cite{hoult_2000}. The effective pulse length,
$\tau$, is chosen such that
$\omega_q\tau=\pi/2$ when $|H_{q}^{\text{tot}}(x)+H_{q}^{\text{tot}}(y)|=1$. 
This procedure specifies the condition that the signal is maximized when the RF field is
unaffected by the sample.

To account for the full volume of metal in which the nuclear spins are 
excited, the areas of each triangle in the surface mesh, $A_q$, as well as skin 
effects must be considered. The latter are incorporated by calculating  the decay of the 
RF signal as a function of the depth, $\gamma$, into the conducting metal surface, leading to
\begin{equation}
  \omega_q(\gamma)=\omega_q(0)\exp\left(-\frac{\gamma}{\delta}\right),
  \label{eq:skin-depth}
\end{equation}
where a skin depth $\delta=10.4$~$\mu$m was employed to match the experimental
conditions used here.

By combining Eqs.~\eqref{eq:trisignal} and \eqref{eq:skin-depth},
an expression for the full MR signal is obtained:
\begin{equation}
  F_q=A_q\int^{\infty}_0
  \omega_q(\gamma)\sin\left(\omega_q(\gamma)\tau\right)d\gamma.
\label{eq:integral}
\end{equation}
which evaluates exactly to,
\begin{equation}
  F_q=\frac{A_q\delta}{\tau}\left[
  1-\cos\left(\omega_q(0)\tau\right)\right]
\label{eq:numintegral}
\end{equation}
This integral was evaluated numerically for each surface triangle.
To account for the overlap of the regions near the edges of the metal, the integrals
for each triangle were truncated at $\gamma=r_{\text{edge}}$, where $r_{\text{edge}}$
is the distance from the center of each triangle to the closest edge of the electrode.

To mimic the imaging experiments, a 2D histogram was built and the $F_q$
values binned depending on their spatial locations relative to the face of
the metal.

\subsubsection{Susceptibility Calculations}

Magnetic field maps were calculated using the FFT method described in 
Refs~\citenum{salomir_fast_2003} and \citenum{marques_2005}.
The method takes as input a 3D grid of susceptibility values
representing the system to be modeled and uses 3D Fourier transforms
of this distribution to efficiently calculate the modifications 
that are made to the local magnetic field at each position due to
susceptibility effects. 
The simulation cell consisted of 512$^{3}$ points representing a cube with
12.8~mm sides. A cuboid in the middle of the cell measuring
3~x~0.375~x~10~mm (in the \emph{x}, \emph{y}, and \emph{z} directions, respectively,
with $ $$B_{0}$ aligned along \emph{z}) corresponded to the lithium metal, and
was assigned a volume susceptibility of 
$\chi^{\text{Li}}_\text{vol}=24.1\times10^{-6}$ \cite{gugan_1997,hedgcock_1960}
in SI units, while the rest of the cell 
was modeled as a vacuum with $\chi_{\text{vol}}=0$. 

The calculation yields the susceptibility-corrected field at each point in the
simulation cell, $H_0^{\text{eff}}$. To match the experimental chemical shifts,
the Knight shift must also be accounted for.  Its contribution is defined as a
constant offset, $K$, that directly scales the resonance frequency,
$H_0^\text{exp}=(1+K)H_0^{\text{eff}}$, where $K=0.0261\%$ for \li7{}
metal \cite{gutowsky_1952,rowland1961nuclear}.

The NMR spectrum was reproduced from the susceptibility results by generating a
separate histogram of the chemical shifts for the surface sites on each side of 
the metal. Artificial line broadening of 2~ppm was used to smoothen
discretization artifacts and to aid the comparison with the experimental spectra.

\section{Results }

\begin{figure}
\begin{centering}
\includegraphics[width=13cm]{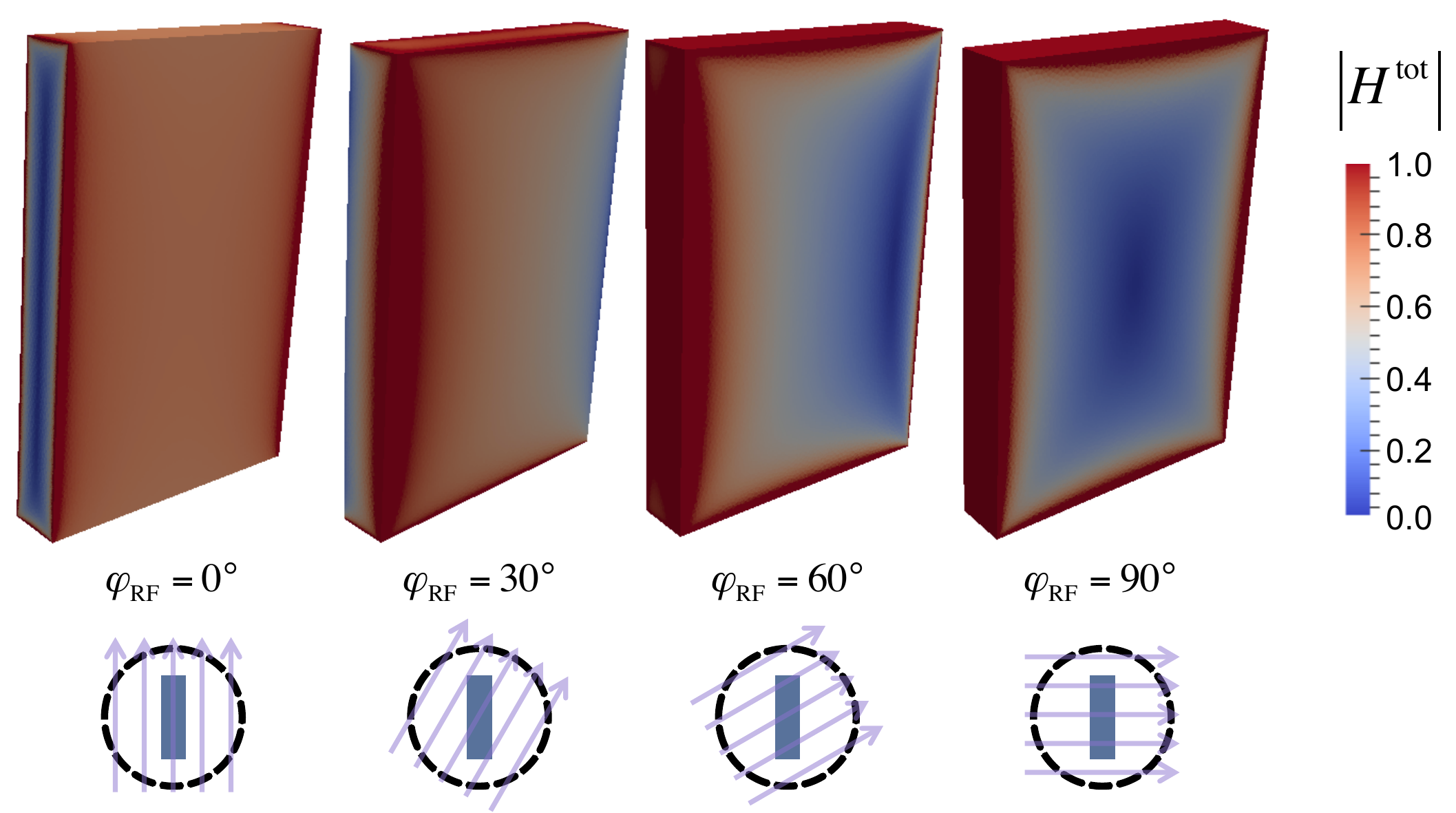}
\par\end{centering}
\caption{\label{fig:field_plot} Plots of the calculated field at the surface
of the metal when the direction of the applied field is rotated
with respect to the metal strip face at \thetarf{}$=0,~30,~60,~90^{\circ}$.
Bottom: Depiction of the relative orientations of the metal strip face and the
direction of field propagation.  }
\end{figure}

Numerical calculations were performed on a Li metal slab at different
relative orientations to the RF field in order 
to explore the angular dependence of the
total RF field strength. 
Plots of the numerically calculated total field, \htot{}
at locations on the surface of the Li metal (Figure~\ref{fig:field_plot})
show that it has a strong dependence on \thetarf{}.
As a general rule, the faces of the metal that are
perpendicular to the propagation direction of the RF field experience a
significantly diminished \htot{} field. 
This reduction can be thought of as being caused by the 
induced eddy currents, which, according
to Lenz's law, produce a magnetic field opposing that of the RF field.
As the eddy currents circulate perpendicular to the the RF field direction,
their effect is much greater on the faces of the metal that are also 
perpendicular to the RF field direction. 

\begin{figure}
\begin{centering}
\includegraphics[width=13cm]{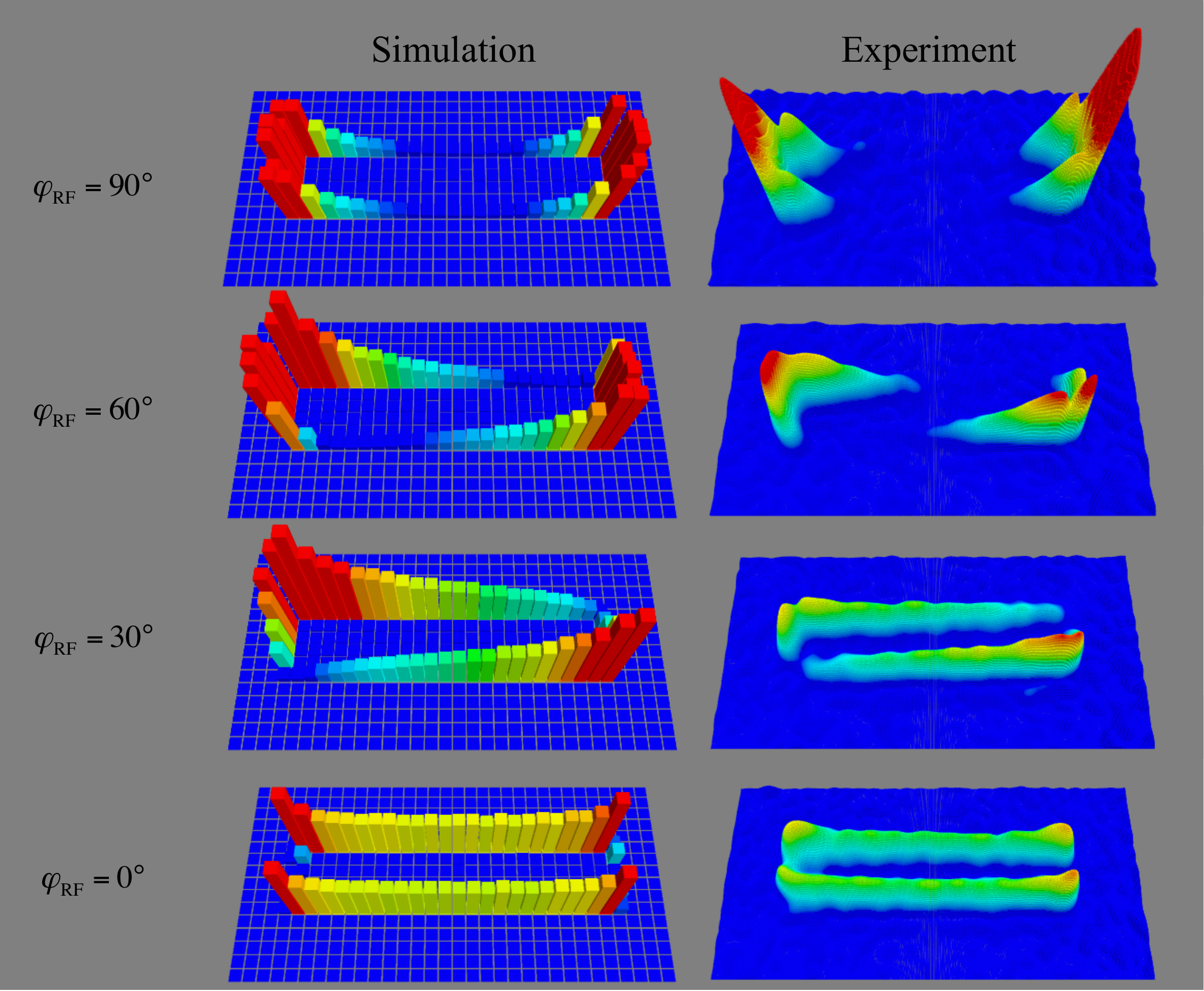}
\par\end{centering}
\caption{\label{fig:mri-results}Results from simulations (left) and experiment
(right) corresponding, from bottom to top, to \thetarf{}$=0,30,60,90^{\circ}$.
The plots within each series are drawn at the same scale. Each experimental
image displays a region of approximately 4.5~x~1.5~mm in the directions parallel
and perpendicular to the major face of the metal, respectively.}
\end{figure}

The spatial variations in the field, displayed in Figure~\ref{fig:field_plot},
give rise to
the  variations in the detected MR signal from the metal surface.
As detailed in the methods section, the MR signal can be simulated from the
$x$ and $y$ components of the 
\htot{} field map and converted into a histogram that shows the \emph{xy}
distribution of MR signal, facilitating a direct comparison with 2D MRI images
(Figure~\ref{fig:mri-results}).
The experimental images are in remarkable agreement with the simulated results
and both inherit recognizable features from the
field maps in Figure~\ref{fig:field_plot}.
At \thetarfn[0], the major, \emph{ac} faces of the metal strip are uniformly
excited, while the minor, \emph{bc} faces show little or no excitation
(see Figure~\ref{fig:electrode} for edge/face labels).
As the field is rotated the signal increases at the \emph{bc} faces until, at
\thetarfn[90], it becomes the main contribution to the overall signal. 
The signal due to the \emph{ab} faces is too small and distributed over a large 
area so is not immediately apparent in the MR images or the calculation results.

There are minor discrepancies between the simulated and experimental results at
the corners of the metal strip when \thetarfn[60] and $90^{\circ}$.
Such discrepancies are expected, because the corners represent numerically challenging 
regions, and the metal plate used in the experiment does not have perfectly sharp edges. 
There is also poor agreement in the relative intensities of the simulated images 
for \thetarfn[60] and $90^{\circ}$, which could possibly be due to an issue with the 
exact pulse calibration.

The excellent overall agreement between the simulated images and those obtained 
experimentally validates the calculation method and the approximations used.
It also confirms the importance of aligning the sample to \thetarfn[0]
in order to acquire artifact-free MR images of bulk metals. 
More significantly, the spatial variation in \htot{} will also enable the use of
nutation experiments that can actively select different faces of a metal strip
or, facilitated by numerical calculations, specific locations on the surface of 
any arbitrarily shaped conductor (on distance scales much larger than the skin-depth). This approach has
great potential in application to battery systems, where it may be possible to
use nutation experiments to select only the active faces of a metal electrode,
giving some degree of spatial selectivity without requiring time-intensive imaging techniques. 

\begin{figure}
\begin{centering}
\includegraphics[width=12cm]{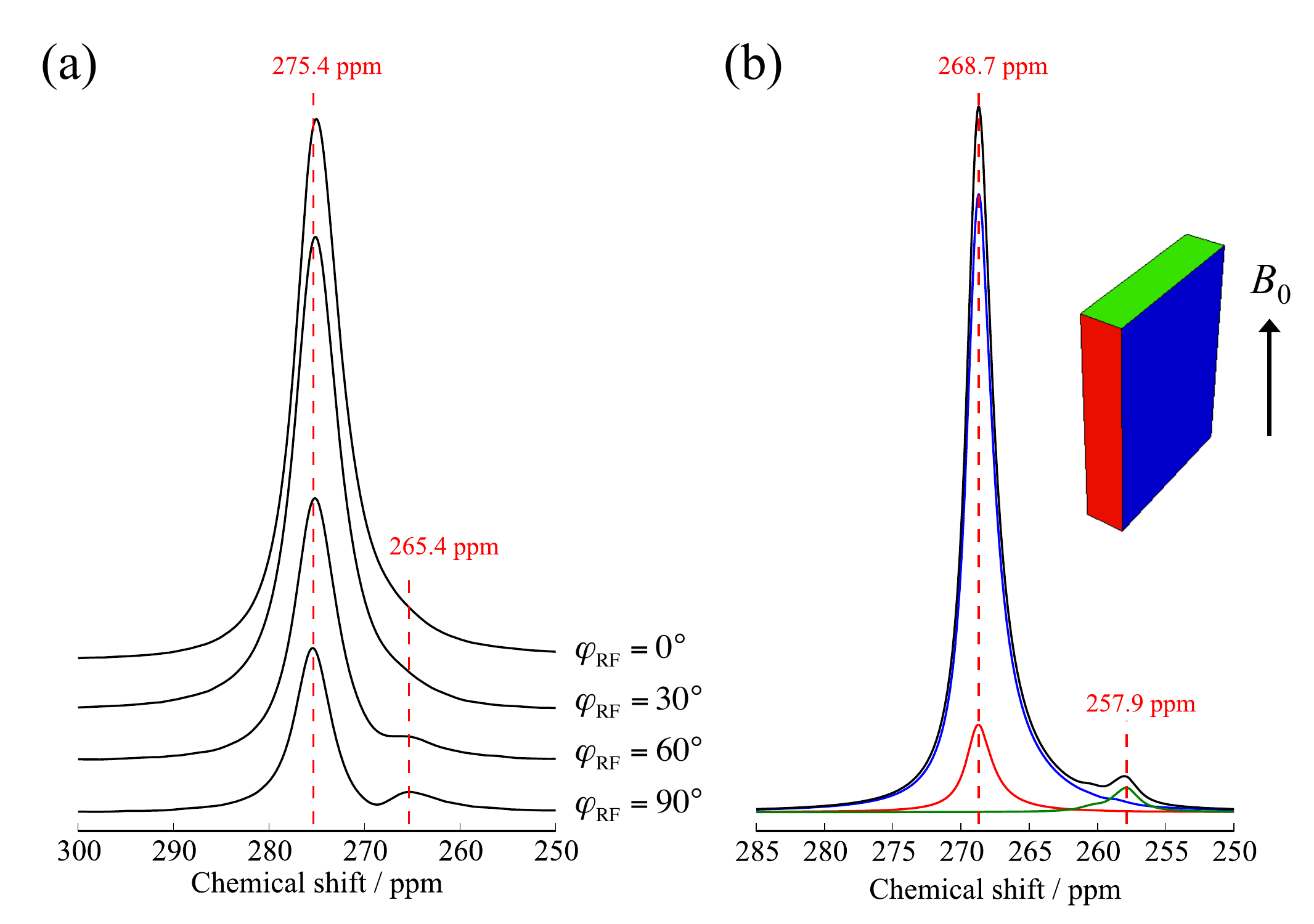}
\par\end{centering}
\caption{\label{fig:1d-spectra}(a) Experimental NMR spectra of the Li-metal
phantom at each orientation, (b) simulated NMR spectrum from the susceptibility
calculations, with line colors corresponding to the faces illustrated
in the accompanying schematic, while the black line is the sum.}
\end{figure}

The dependence on the relative RF field orientation also has far reaching
consequences for NMR spectroscopy experiments performed on samples that are good conductors,
and particularly for the quantification of the NMR signal.
As expected, 1D NMR spectra
of the \li7{} metal strip at different field orientations
(Figure~\ref{fig:1d-spectra}(a)), show a strong dependence of the total signal
intensity on \thetarf{}.
These changes follow directly from the results of Figure~\ref{fig:mri-results},
but the relative signal contributions will also depend on sample orientation.
One clue that this is indeed 
observed, is in the apparent emergence of the peak at $\approx$265~ppm  
when \thetarfn[90]. The susceptibility calculations 
(Figure~\ref{fig:1d-spectra}(b)) show that the $^7$Li nuclei on the top and 
bottom (\emph{ab}, in green) faces of the metal are shifted upfield with respect 
to the nuclei on the other faces, solely due to susceptibility effects. 
Meanwhile, the field calculations show that \htot{}
on the \emph{ab} faces has a minimal dependence on \thetarf{} 
(increasing only slightly from \thetarfn[0] to \thetarfn[90]) and so
the overall signal from these faces should stay constant with \thetarf{}, 
while the contributions from the other faces vary strongly. 
Indeed, this is precisely what is observed in the experimental spectra;
the intensity of the upfield peak is almost unchanged 
throughout the series 
(it is still observed as a shoulder on the larger peak at \thetarfn[0])
while the intensity of the peak at $\approx$275~ppm 
reduces on moving from \thetarfn[0] to \thetarfn[90].

It is noted that the shifts of the calculated peaks
in Figure~\ref{fig:1d-spectra}(b) do not match exactly with the experimental
values in Figure~\ref{fig:1d-spectra}(a), although the separation of the peaks,
$\approx$10~ppm, is reproduced. 
The calculation depends only on the constants for the susceptibility and 
Knight shift of Li metal, but assumes a perfect cuboid of the metal 
($\chi=\chi^{\text{Li}}_{\text{vol}}$)
in a vacuum ($\chi=0$). 
There is some variation in the literature regarding the exact values of 
$K$ and $\chi^{\text{Li}}_{\text{vol}}$ which may contribute to the 
discrepancy in the shifts. 
Any diamagnetic chemical impurities on 
the surface of the Li metal, formed due to reactions with air, or leftover
from the preparation technique (the surface of the metal can retain traces of a
residue left over from manufacturing, some of which may remain even after cleaning),
will also accentuate the shift difference between the faces and impact the shifts. 

Despite the slight differences in the chemical shift values, the behavior of the
extra peak upon changes to \thetarf{}, and its upfield shift, support its
assignment to Li sites on the \emph{ab} faces of the metal. Similar assignments 
have also been made in battery systems, where Li-microstructures growing 
perpendicular to the surface of the electrode have been shown to display a
significant shift away from the bulk metal peak \cite{bhattacharyya_situ_2010,
chandrashekar_7li_2012}, and where orientation-dependent bulk magnetic
susceptibility effects have been explored \cite{zhou_2013}. In non-conducting
samples, the latter leads to a broadening of the spectrum as the shifts vary
continuously from one part of the sample to another.  In conductors, however,
the skin depth effects ensure that only surface sites are probed, hence discrete
sets of shifts can be observed without the associated broadening. It is possible
that by manipulating the shape of the electrodes, controlling \thetarf{}, and
explicitly considering the materials surrounding the electrodes, the resolution
between these sites could be increased so that certain faces of an electrode
could be preferentially probed using MR techniques.  The susceptibility
calculations used here \cite{salomir_fast_2003} are fully flexible and could be
adapted to aid in the design of such systems. 

\section{Conclusions}

The dependence of the RF field on the orientation of bulk metal plates placed within
the RF coil has been investigated both experimentally and theoretically. 
 This dependence was
explored using numerical field calculations that have been validated by
direct comparison to experimental MRI results.  The results show that not all of the
faces of a conducting cuboid can be uniformly excited at any orientation,
with the faces whose normals are parallel to the field direction experiencing
greatly reduced fields. Furthermore, edges and corners 
typically produce larger signals in any orientation.
Given the calculated RF fields and susceptibility shifts, 
the NMR spectra of a metal plate at any orientation can be fully reconstructed.
It is envisaged that the insights into the spatial variations of the field
offered by the calculations will enable quantitative surface NMR and MRI studies of 
bulk metals. Furthermore, through a combination of sample shape and orientation, 
studies directed at  diagnosing surface roughness or microstructure growth could be performed. 
This approach should
be particularly useful for NMR and MRI of battery systems or other electrochemical devices.
In combination with susceptibility calculations, this procedure allows the
NMR spectra of conductors to be fully described.

\section*{Acknowledgements}

This work is part of the NECCES (Northeastern Center for Chemical Energy Storage), 
an Energy Frontier
Research Center funded by the U.S. Department of Energy (DOE), Office 
of Basic Energy Sciences under Award Number
DE-SC0001294 (in situ methodology), and was funded by matching support from 
NYSTAR-NYSDED, and by the New York State Energy Research Development
Authority (NYSERDA) (H.J.C.).
The work of L.G. and A.K. was supported by the Applied Mathematical Sciences 
Program of the U.S. Department of Energy under Contract DEFGO288ER25053 and 
by the Office of the Assistant Secretary of Defense for Research and 
Engineering and AFOSR under NSSEFF Program Award FA9550-10-1-0180.

\section*{References}

\bibliographystyle{elsarticle-num}
\bibliography{num-method,Li-rotations}

\end{document}